\documentclass[nofootinbib]{revtex4}
\usepackage[dvips]{color}
\usepackage{graphicx}
\usepackage{epsfig}

\begin{document}

\title{Non-adiabatic perturbations in decaying vacuum cosmology}

\author{W. Zimdahl$^{1}$\email{winfried.zimdahl@pq.cnpq.br}, H. A. Borges$^{2,3}$\email{humberto@ufba.br}, S. Carneiro$^{2}$\footnote{ICTP Associate Member.}\email{saulo.carneiro@pq.cnpq.br}, J. C. Fabris$^1$\email{fabris@pq.cnpq.br}, W. S. Hip\'{o}lito-Ricaldi$^{4}$\email{whipolito@gmail.com}}

\affiliation{$^1$Departamento de F\'{\i}sica, Universidade Federal do Esp\'{\i}rito Santo, Vit\'oria, ES, Brazil\\ $^2$Instituto de F\'{\i}sica, Universidade Federal da Bahia, Salvador, BA, Brazil\\ $^3$CFP, Universidade Federal do Rec\^oncavo da Bahia, Amargosa, BA, Brazil\\ $^4$Departamento de Ci\^encias Matem\'aticas e Naturais, UFES, S\~{a}o Mateus, ES, Brazil}

\date{\today}

\begin{abstract}

We investigate a spatially flat Friedmann-Lema\^{i}tre-Robertson-Walker cosmology in which a decaying vacuum term causes matter production at late times. Assuming a decay proportional to the Hubble rate, the ratio of the background energy densities of dark matter and dark energy changes with the cosmic scale factor as $a^{-3/2}$.
The intrinsically non-adiabatic two-component perturbation dynamics of this model is reduced to a single second-order equation. Perturbations of the vacuum term are shown to be negligible on scales that are relevant for structure formation. On larger scales, dark-energy perturbations give a somewhat higher contribution but remain always smaller than the dark-matter perturbations.

\end{abstract}

\maketitle

\section{Introduction}

The origin of the accelerated expansion of the Universe, first described in \cite{Riess}, is certainly the most challenging problem of modern cosmology, receiving considerable attention today in both theoretical and observational areas \cite{rev}. The simplest explanation is a cosmological constant $\Lambda$ in Einstein's equations which remained the favored option until today and led to the
$\Lambda$CDM model which also plays the role of a reference model for most studies in the field.
Such a constant is usually associated to the energy density of quantum vacuum fluctuations. Because of the cosmological constant problem in its different facets, including the coincidence problem, a great deal of work was devoted to alternative approaches in which a similar dynamics as that of the $\Lambda$CDM model is reproduced with a time varying cosmological term, i.e., the cosmological constant is dynamiz ed.
So far, the vacuum energy is not well determined by quantum field theories in curved space-time, given the divergences involved in its derivation. For the well-established renormalization method  in the case of free conformal fields in an early de Sitter space-time \cite{H4,award} the resulting energy density is proportional to $H^4$, where $H$ is the Hubble parameter.  But its value today is very small as compared to the observed $\Lambda$. A correct order of magnitude is obtained for a dependence $\Lambda \sim H^2$ \cite{cohen}. This circumstance gave rise to holographic dark-energy models (see, e.g. \cite{WD}, and references therein). Further semi-phenomenological ans\"{a}tze for a time-dependent cosmology term can be found in the literature \cite{Lambda(t)}.
Among them there is the proposal that the QCD vacuum condensate associated to the chiral phase transition leads to a vacuum density proportional to $H$ \cite{QCD}, more precisely $\Lambda = m^3 H$, where $m \approx 150 $MeV is the energy scale of the QCD phase transition.
It is this approach that we are interested in in the present paper.
The corresponding cosmological model is qualitatively similar to the standard one (the $\Lambda$CDM model), with a radiation phase followed by a phase dominated by matter and, subsequently, by the cosmological term \cite{Borges}. The important point is that the time variation of the cosmological term is concomitant with a process of matter production in order to assure the conservation of the total energy. In this sense, the model may be seen as a particular case of interacting dark-energy models \cite{interacting}. The $\Lambda \propto H$ model has been tested from the observational viewpoint at both the background and perturbation levels. A preliminary joint analysis of the redshift-distance relation for supernovas of type Ia, the baryonic acoustic oscillations and the position of the first peak in the anisotropy spectrum of the cosmic microwave background  was performed in \cite{anterior}, resulting in a present matter density parameter $\Omega_{M0} \approx 0.36$.
At the perturbation level, the matter power spectrum has been calculated by assuming that the interacting cosmological term is strictly homogeneous \cite{Julio}. This spectrum reproduced the data for an even higher matter density parameter, $\Omega_{M0} \approx 0.48$.


In the present paper we try to clarify the situation by reconsidering the previous analysis in a gauge-invariant setting. This allows us to understand the role of non-adiabatic perturbations in decaying vacuum cosmology. For simplicity we restrict ourselves to the  dynamically most relevant components, namely dark matter and dark energy, i.e., we neglect both the baryon and the radiation components.
We demonstrate that the entire two-component perturbation dynamics can be reduced to a single second-order perturbation equation for the density contrast of non-relativistic matter.
We show also that at high redshifts the non-adiabaticity of the model is small, although non-vanishing, which allows us to use adiabatic initial conditions as a good approximation.
In particular, we address the problem of the relevance of perturbations of the dark energy component for structure formation. In many investigations it is assumed from the outset that dark energy does not cluster on small scales. But strictly speaking, this has to be justified on a case-by-case basis for all dynamical dark energy models.
As it was argued by Park et al. \cite{Park-Hwang}, neglecting the perturbations of the dark energy component may lead to inconsistencies and unreliable conclusions concerning the interpretation of observational data.
For the present model of a decaying cosmological term, however, we shall show explicitly that the dark-energy fluctuations are indeed smaller than the dark-matter fluctuations by several orders of magnitude on scales that are relevant for structure formation.
This justifies \textit{a posteriori} the already mentioned previous analysis in which fluctuations of the dark energy component were not taken into account \cite{Julio} and which was compatible with a value of $\Omega_{M0} \approx 0.48$ for the matter density parameter. This value is considerably higher than the corresponding $\Lambda$CDM-value and also higher than the background concordance value obtained in \cite{anterior}. However, an update of the background tests with the most recent surveys of type Ia-supernovae seems to admit matter density parameter up to this value, at least for some of the data sets \cite{JailsonIII}.
On superhorizon scales the fluctuations of the dark energy contribute a larger fraction to the total energy density perturbation but still remain smaller than the matter fluctuations.

The paper is organized as follows. In section \ref{A two-component fluid} we introduce the general expressions for our interacting two-component fluid, obtain the homogeneous and isotropic background dynamics of the model and perform a detailed analysis of the interaction term. In Section \ref{Perturbation dynamics} we derive the central second-order equation for the matter perturbations in terms of which we also determine the perturbations of the dark-energy component. Furthermore, we discuss the non-adiabatic nature of the fluctuations and show that the non-adiabaticity is negligible at early times. Section \ref{The power spectrum} presents the results of a numerical analysis which quantifies the role of perturbations of the dark-energy component.  Finally, section \ref{Conclusions} is devoted to conclusions and remarks.

\section{The two-component fluid model}
\label{A two-component fluid}

\subsection{Basic properties}

We consider a two-component system with a total energy momentum
tensor
\begin{equation}
T_{ik} = \rho u_{i}u_{k} + p h_{ik}\ , \qquad T_{\ ;k}^{ik} = 0,\
\label{T}
\end{equation}
where $h _{ik}=g_{ik} + u_{i}u_{k}$ and $g_{ik}u^{i}u^{k} = -1$. The quantity $u^{i}$ denotes the total four-velocity of the cosmic substratum. Latin indices run from $0$ to $3$.
We assume a split of $T_{ik}$ into a matter component (subindex M) and a dark energy component (subindex X),
\begin{equation}\label{Ttot}
T^{ik} = T_{M}^{ik} + T_{X}^{ik},
\end{equation}
with ($A= M, X$)
\begin{equation}\label{TA}
T_{A}^{ik} = \rho_{A} u_A^{i} u^{k}_{A} + p_{A} h_{A}^{ik} \
,\qquad\ h_{A}^{ik} = g^{ik} + u_A^{i} u^{k}_{A} \ .
\end{equation}
Furthermore, we admit an interaction between the components:
\begin{equation}\label{Q}
T_{M\ ;k}^{ik} = Q^{i},\qquad T_{X\ ;k}^{ik} = - Q^{i}\ .
\end{equation}
Then, the separate energy-balance equations are
\begin{equation}
-u_{Mi}T^{ik}_{M\ ;k} = \rho_{M,a}u_{M}^{a} +  \Theta_{M} \left(\rho_{M} + p_{M}\right) = -u_{Ma}Q^{a}\
\label{eb1}
\end{equation}
and
\begin{equation}
-u_{Xi}T^{ik}_{X\ ;k} = \rho_{X,a}u_{X}^{a} +  \Theta_{X} \left(\rho_{X} + p_{X}\right) = u_{Xa}Q^{a}\ .
\label{eb2}
\end{equation}
In general, each component has its own four-velocity, with $g_{ik}u_{A}^{i}u_{A}^{k} = -1$. The quantities $\Theta_{A}$ are defined as $\Theta_{A} = u^{a}_{A;a}$. For the homogeneous and isotropic background we assume $u_{M}^{a} = u_{X}^{a} = u^{a}$. Likewise, we have the momentum balances
\begin{equation}
h_{Mi}^{a}T^{ik}_{M\ ;k} = \left(\rho_{M} + p_{M}\right)\dot{u}_{M}^{a} + p_{M,i}h_{M}^{ai} = h_{M i}^{a} Q^{i}\
\label{mb1}
\end{equation}
and
\begin{equation}
h_{Xi}^{a}T^{ik}_{X\ ;k} = \left(\rho_{X} + p_{X}\right)\dot{u}_{X}^{a} + p_{X,i}h_{X}^{ai} = - h_{X i}^{a} Q^{i},\
\label{mb2}
\end{equation}
where $\dot{u}_{A}^{a} \equiv u_{A ;b}^{a}u_{A}^{b}$.
The source term $Q^{i}$ is split into parts proportional and perpendicular to the total four-velocity according to
\begin{equation}
Q^{i} = u^{i}Q + \bar{Q}^{i}\ ,
\label{Qdec}
\end{equation}
where $Q = - u_{i}Q^{i}$ and $\bar{Q}^{i} = h^{i}_{a}Q^{a}$, with $u_{i}\bar{Q}^{i} = 0$. (Alternatively, one could have introduced a split with respect to the matter four-velocity. As we shall see later, for the model of interest here both options lead to identical results.)

The contribution $T_{X}^{ik}$ is supposed to describe some form of dark energy. In the simple case of an equation of state $p_{X} = - \rho_{X}$, where $\rho_{X}$ is not necessarily constant, we have
\begin{equation}
T_{X}^{ik} = - \rho_{X}g^{ik}\ .
\label{Tx}
\end{equation}
We are interested here in the case
\begin{equation}
\rho_{X} = \frac{\sigma}{3}\Theta\ ,
\label{rX}
\end{equation}
where $\Theta \equiv u^{a}_{;a}$ is the expansion scalar and  $\sigma$ is a constant. In the homogeneous and isotropic background one has $\Theta = 3H$, where $H$ is the Hubble rate.

\subsection{Background dynamics}
\label{Background dynamics}

In the homogeneous and isotropic background we have
\begin{equation}
3 H^{2} = 8\pi G \rho = 8\pi G \left(\rho_{M} + \rho_{X}\right) =  8\pi G \left(\rho_{M} + \sigma H\right)\
\label{fried}
\end{equation}
and, assuming $p_{M}=0$ from now on,
\begin{equation}
\dot{H} = - 4\pi G  \left(\rho + p\right) = - 4\pi G \rho_{M} \ .
\label{dH}
\end{equation}
Combining (\ref{fried}) and (\ref{dH}) we obtain
\begin{equation}
\dot{H} = - \frac{3}{2}H^{2} +  4\pi G \sigma H\ .
\label{dH1}
\end{equation}
With
\begin{equation}
3H_{0}^{2} = 8\pi G \rho_{0}\ ,\quad \Omega_{M0}\equiv \frac{\rho_{M0}}{\rho_{0}}\quad \mathrm{and} \quad\sigma = \frac{\rho_{0}}{H_{0}}\left(1 - \Omega_{M0}\right)\ ,
\label{sig}
\end{equation}
integration yields the Hubble rate \cite{Borges}
\begin{equation}
H = H_{0}\left[1 - \Omega_{M0} + \Omega_{M0}a^{-3/2}\right]
\ ,
\label{Hom}
\end{equation}
where a subindex 0 indicates the present value of the corresponding variable. The quantity $a$ denotes the scale factor of the Robertson-Walker metric which was normalized to $a_{0}=1$.
The energy densities are given by \cite{Borges}
\begin{equation}
\frac{\rho_{M}}{\rho_{0}} = \Omega_{M0}a^{-3/2}\left[1 - \Omega_{M0} + \Omega_{M0}a^{-3/2}\right]
\
\label{rmom}
\end{equation}
and
\begin{equation}
\frac{\rho_{X}}{\rho_{0}} = \left(1 - \Omega_{M0}\right)\left[1 - \Omega_{M0} + \Omega_{M0}a^{-3/2}\right]
\ ,
\label{rxom}
\end{equation}
respectively. For $\sigma =0$ as well as for $a<<1$ we consistently recover the Einstein - de Sitter universe. For $a>>1$, the solution tends to de Sitter space-time.
The ratio of the energy densities is
\begin{equation}
\frac{\rho_{M}}{\rho_{X}} = \frac{\Omega_{M0}}{1 - \Omega_{M0}}a^{-3/2}\ .
\label{ratioom}
\end{equation}
It decays with $a^{-3/2}$, i.e., with a lesser rate than in the $\Lambda$CDM model for which the corresponding ratio decays
as $a^{-3}$. But in both cases it approaches zero in the long-time limit.

The balances (\ref{eb1}) and (\ref{eb2}) take the forms
\begin{equation}
\dot{\rho}_{M} + 3H\rho_{M} = Q^{0}
\label{eb1+}
\end{equation}
and
\begin{equation}
\dot{\rho}_{X}  = - Q^{0}\ ,
\label{eb2+}
\end{equation}
respectively. Since $\rho_{X}$ is given as $\rho_{X} = \frac{\sigma}{3}\Theta = \sigma H$, the background source terms are
\begin{equation}
Q^{0} = u^{0}Q  = Q = \dot{p}_{X} = - \frac{\sigma}{3}\dot{\Theta} = - \sigma\dot{H}\quad \mathrm{and} \quad Q^{\alpha} = 0\ .
\label{Q0}
\end{equation}
Alternatively,
\begin{equation}
u_{a}Q^{a} = - Q = -\dot{p}_{X} =  \frac{\sigma}{3}\dot{\Theta} =  \sigma\dot{H}\quad \mathrm{and} \quad \bar{Q}^{a} = 0\ .
\label{Q01}
\end{equation}
Together with (\ref{dH})
we find the  balance equations
\begin{equation}
\dot{\rho}_{M} + 3 H \rho_{M} = 4\pi G \sigma\rho_{M}\
\label{drmq}
\end{equation}
and
\begin{equation}
\dot{\rho}_{X}  = -4\pi G \sigma\rho_{M}\ .
\label{drxq}
\end{equation}
Differentiating the expressions (\ref{rmom}) and (\ref{rxom}) one realizes that they indeed are solutions of
(\ref{drmq}) and (\ref{drxq}).

\subsection{The perturbed source term}
\label{The perturbed source term}

Denoting first-order perturbations by a hat symbol and recalling that for the background $u_{M}^{a} = u_{X}^{a} = u^{a}$ is valid, the perturbed time components of the four-velocities are
\begin{equation}
\hat{u}_{0} = \hat{u}^{0} = \hat{u}_{M}^{0} =\hat{u}_{X}^{0}  = \frac{1}{2}\hat{g}_{00}\ .
\label{u0}
\end{equation}
According to the perfect-fluid structure of both the total energy-momentum tensor (\ref{T}) and the energy-momentum tensors of the components in (\ref{TA}), and with $u_{M}^{a} = u_{X}^{a} = u^{a}$ in the background, we have first-order energy-density perturbations
$\hat{\rho} = \hat{\rho}_{M} + \hat{\rho}_{X}$, pressure perturbations $\hat{p} = \hat{p}_{M} + \hat{p}_{X} = \hat{p}_{X}$
and
\begin{equation}
\hat{T}^{0}_{\alpha} = \hat{T}^{0}_{M\alpha} + \hat{T}^{0}_{X\alpha}\quad\Rightarrow\quad
\left(\rho + p\right)\hat{u}_{\alpha} = \rho_{M}\hat{u}_{M\alpha} + \left(\rho_{X} + p_{X}\right)\hat{u}_{X\alpha}
\ .
\label{T0al}
\end{equation}
Greek indices run from $1$ to $3$. For $p_{X} = - \rho_{X}$ it follows
\begin{equation}
p_{X} = - \rho_{X} \quad\Rightarrow\quad \rho + p = \rho_{M} \quad\Rightarrow\quad\hat{u}_{M\alpha} = \hat{u}_{\alpha}\ .
\label{ual}
\end{equation}
Since the component $M$ is supposed to describe matter, it is clear from (\ref{T0al}) that the perturbed matter velocity $\hat{u}_{M\alpha}$ coincides with the total velocity perturbation $\hat{u}_{\alpha}$.
With $u^{n}_{M} = u^{n}$ up to first order, the energy balance in (\ref{eb1}) (correct up to first order) can be written as
\begin{equation}
\rho_{M,a}u^{a} = -  \Theta \rho_{M}  -u_{a}Q^{a}\ .
\label{eb1u}
\end{equation}
On the other hand, the total energy balance is
\begin{equation}
\rho_{,a}u^{a} = - \Theta \left(\rho + p\right)
\ .
\label{eb}
\end{equation}
For the difference it follows that
\begin{equation}
\dot{\rho} - \dot{\rho}_{M} \equiv \left(\rho - \rho_{M} \right)_{,a}u^{a}
= u_{a}Q^{a}
\ .
\label{diffeb}
\end{equation}
Since, at least up to linear order, $\rho - \rho_{M} = \rho_{X}$, equation (\ref{diffeb}) is equivalent (up to the first order) to
\begin{equation}
\dot{\rho}_{X} \equiv \rho_{X,a}u^{a} = u_{a}Q^{a}
\ .
\label{drx}
\end{equation}
In zeroth order we recover (\ref{eb2+}) with (\ref{Q0}).
The first-order equation is (cf. (\ref{u0}))
\begin{equation}
\dot{\hat{\rho}}_{X} + \dot{\rho}_{X}\hat{u}^{0} = \left(u_{a}Q^{a}\right)^{\hat{}}
\ .
\label{diffepert}
\end{equation}
Notice that (\ref{diffepert}) results from a combination of the total energy conservation and the matter energy balance. It has to be consistent with the dark energy balance  (\ref{eb2}). At first order, the latter becomes
\begin{equation}
\dot{\hat{\rho}}_{X} + \dot{\rho}_{X}\hat{u}^{0} = \left(u_{Xa}Q^{a}\right)^{\hat{}}
\ .
\label{eb2pert}
\end{equation}
This means that
\begin{equation}
\left(u_{Xa}Q^{a}\right)^{\hat{}} = \left(u_{a}Q^{a}\right)^{\hat{}}
\ ,
\label{u2u}
\end{equation}
i.e., the projections of $Q^{a}$ along $u_{Xa}$ and along $u_{a}$ coincide. Explicitly,
\begin{equation}
\left(u_{a}Q^{a}\right)^{\hat{}} = \left(u_{a}u^{a}Q\right)^{\hat{}} = - \hat{Q} = \frac{\sigma}{3}\hat{\dot{\Theta}}
\ .
\label{uQpert}
\end{equation}
In a next step we consider the momentum balances. The total momentum conservation is described by
\begin{equation}
h_{i}^{a}T^{ik}_{\ ;k} = \left(\rho_{M} + \rho_{X} + p_{X}\right)\dot{u}^{a} + h_{}^{ai} p_{X, i} = 0\ .
\label{}
\end{equation}
With $p_{X} = - \rho_{X}$ and $p_{X} = - \frac{\sigma}{3}\Theta$, we have
\begin{equation}
h_{i}^{a}T^{ik}_{\ ;k} = \rho_{M}\dot{u}^{a} - h^{ai}\frac{\sigma}{3}\Theta_{, i} = 0\ .
\label{mb}
\end{equation}
Using $u^{n}_{M} = u^{n}$ again, the momentum balance (\ref{mb1}) for the matter component  becomes
\begin{equation}
h_{i}^{a}T^{ik}_{M\ ;k} = \rho_{M}\dot{u}^{a} = h^{ai}Q_{i}\ .
\label{mb1+}
\end{equation}
Consistency between (\ref{mb}) and (\ref{mb1+}) requires
\begin{equation}
\bar{Q}^{a} \equiv h^{ai}Q_{i} = \frac{\sigma}{3}h^{ai}\Theta_{,i} \ .
\label{hQ}
\end{equation}
Notice that we have only used the total momentum conservation and the matter momentum balance.
The momentum balance (\ref{mb2}) of the dark energy degenerates for the case $p_{X} = - \rho_{X}$. It does not describe any dynamics.

With (\ref{hQ}), the source term in the momentum balance is explicitly known. Up to first order the matter momentum balance (\ref{mb1+}) reduces to
\begin{equation}
\rho_{M}\dot{u}_{a} = \bar{Q}_{a}
\ .
\label{BMb}
\end{equation}
Explicitly, from (\ref{hQ}) the first-order source term becomes
\begin{equation}
\hat{\bar{Q}}_{0} = 0\ , \qquad \hat{\bar{Q}}_{\alpha} = \frac{\sigma}{3}\left[\hat{\Theta}_{,\alpha} +\hat{u}_{\alpha}\dot{\Theta}\right]
\ .
\label{bQ}
\end{equation}

\subsection{Basic set of equations}

Before explicitly starting the perturbative analysis, we summarize  the basic set of equations.
It consists of the energy balance
\begin{equation}
\dot{\rho}_{M} + \Theta \rho_{M} = -u_{a}Q^{a} = Q = -\frac{\sigma}{3}\dot{\Theta} \ ,
\label{BE}
\end{equation}
the momentum balance
\begin{equation}
\rho_{M}\dot{u}_{a} = \frac{\sigma}{3}h_{a}^{i}\Theta_{,i}
\ ,
\label{BM}
\end{equation}
and the Raychaudhuri equation
\begin{equation}
\dot{\Theta} + \frac{1}{3}\Theta^{2} - \dot{u}^{a}_{;a} + 4\pi G \left(\rho + 3
p\right) = 0\ .\label{Ray}
\end{equation}
In the present case ($p_{M}=0$ and $p_{X} = - \rho_{X} = - \frac{\sigma}{3}\Theta$) the latter reduces to
\begin{equation}
\dot{\Theta} + \frac{1}{3}\Theta^{2} - \dot{u}^{a}_{;a} + 4\pi G \rho_{M} - \frac{8\pi G}{3}\sigma\Theta = 0\ .\label{Rayred}
\end{equation}
Notice that the source terms in the balances (\ref{BE}) and (\ref{BM}) are given by derivatives (temporal and spatial, respectively) of the expansion scalar.
In the background, the energy balance (\ref{BE}) reduces to (\ref{eb1+}) with (\ref{Q0}), the momentum balance (\ref{BM}) is zero identically and the Raychaudhuri equation (upon using Friedmann's equation) specifies to (\ref{dH1}).
The zeroth-order solutions of the system (\ref{BE}), (\ref{BM}) and (\ref{Rayred}) are given by (\ref{rmom}), (\ref{rxom}) and (\ref{Hom}), respectively.

\section{Perturbation dynamics}
\label{Perturbation dynamics}

\subsection{General relations}
The general line element for scalar perturbations is
\begin{equation}
\mbox{d}s^{2} = - \left(1 + 2 \phi\right)\mbox{d}t^2 + 2 a^2
F_{,\alpha }\mbox{d}t\mbox{d}x^{\alpha} +
a^2\left[\left(1-2\psi\right)\delta _{\alpha \beta} + 2E_{,\alpha
\beta} \right] \mbox{d}x^\alpha\mbox{d}x^\beta \ .\label{ds}
\end{equation}
For the perturbed spatial components of the four-velocity one has
\begin{equation}
a^2\hat{u}^\mu + a^2F_{,\mu} = \hat{u}_\mu \equiv v_{,\mu} \ ,
\label{}
\end{equation}
which defines the velocity perturbation $v$. A choice
$v=0$ corresponds to the comoving gauge. Taking into account the definition $\dot{\rho}_{M} \equiv \rho_{M,a}u^{a}$, the balance (\ref{BE}) becomes in first order
\begin{equation}
\dot{\delta}_{M} + 4\pi G \sigma \delta_{M} - \phi \left(- 3H + 4\pi G\sigma\right) + \hat{\Theta} = \frac{\hat{Q}}{\rho_{M}}\ ,
\label{BE1}
\end{equation}
where we have introduced the first-order fractional perturbation $\delta_{M} \equiv \frac{\hat{\rho}_{M}}{\rho_{M}}$.
With the definition $\dot{u}_{a} \equiv u_{a;b}u^{b}$ which leads to $(\dot{u}_{\alpha})^{\hat{}} = \dot{\hat{u}}_{\alpha} + \phi_{,\alpha}$, the momentum balance (\ref{BM}) becomes in first order
\begin{equation}
\dot{v}_{,\alpha} + \phi_{,\alpha} - \frac{\sigma}{3\rho_{M}}\left(\hat{\Theta} + \dot{\Theta}v\right)_{,\alpha} =0 \quad \Rightarrow\quad \dot{v}
 + \phi = \frac{\sigma}{3\rho_{M}}\left(\hat{\Theta} + \dot{\Theta}v\right)
\ .
\label{BM1}
\end{equation}
To calculate the source term $\hat{Q}$ in (\ref{uQpert}) we have the perturbed Raychaudhuri equation.
With
\begin{equation}
\dot{u}^{a}_{;a} = - \frac{1}{\rho_{M}a^{2}}\left(\Delta\hat{p}_{X} + \dot{p}_{X}\Delta v\right) = \frac{\sigma}{3\rho_{M}a^{2}}\left(\Delta\hat{\Theta} + \dot{\Theta}\Delta v\right)
\ ,
\label{du}
\end{equation}
where $\Delta$ is the three-dimensional Laplacian,
one obtains
\begin{equation}
\hat{\dot{\Theta}} = - \frac{2}{3}\Theta\hat{\Theta} + \frac{\sigma}{3\rho_{M}a^{2}}\left(\Delta\hat{\Theta} + \dot{\Theta}\Delta v\right) - 4\pi G \rho_{M}\delta_{M} + \frac{8\pi G}{3}\sigma \hat{\Theta}
\ .
\label{dT}
\end{equation}
It follows that the source term $\frac{\hat{Q}}{\rho_{M}}$ in (\ref{BE1}) becomes
\begin{equation}
\frac{\hat{Q}}{\rho_{M}} = - \frac{\sigma}{3}\left[\left(\frac{8\pi G}{3}\sigma - \frac{2}{3}\Theta\right)\frac{\hat{\Theta}}{\rho_{M}} + \frac{\sigma}{3\rho_{M}^{2}a^{2}}\left(\Delta\hat{\Theta} + \dot{\Theta}\Delta v\right) - 4\pi G \delta_{M}\right]
\ .
\label{hQ/}
\end{equation}
It is now convenient to describe the dynamics in terms of the gauge-invariant quantities
\begin{equation}
\hat{\Theta}^{c} \equiv \hat{\Theta} + \dot{\Theta} v \ , \quad \delta_{M}^{c} \equiv \delta_{M} + \frac{\dot{\rho}_{M}}{\rho_{M}} v \quad \mathrm{and}\quad \delta_{X}^{c} \equiv \delta_{X} + \frac{\dot{\rho}_{X}}{\rho_{X}} v
\ ,
\label{gi}
\end{equation}
which characterize the perturbations of the expansion scalar, the matter density and the dark energy density, respectively, on comoving hypersurfaces.
Then, equation (\ref{BE1}) takes the form
\begin{equation}
\dot{\delta}_{M}^{c} + 4\pi G \sigma \delta_{M}^{c}  + \left[1 + \frac{\sigma}{3\rho_{M}}\left(3H - 4\pi G\sigma\right)\right]\hat{\Theta}^{c} = \frac{\hat{Q}^{c}}{\rho_{M}}\
\ .
\label{BEM}
\end{equation}
Here, $\hat{Q}^{c} \equiv \hat{Q} + \dot{Q}v$ is the gauge-invariant perturbation of the source term $Q$ with the explicit (momentum space) structure
\begin{equation}
\frac{\hat{Q}^{c}}{\rho_{M}} = - \frac{\sigma}{3}\left[\left(\frac{8\pi G}{3}\sigma - 2 H - \frac{\sigma}{3\rho_{M}}\,\frac{k^{2}}{a^{2}}\right)\frac{\hat{\Theta}^{c}}{\rho_{M}} - 4\pi G \delta_{M}^{c}\right]
\ ,
\label{hQck/}
\end{equation}
where $k^{2}$ is the square of the comoving wave vector.
Eq.~(\ref{BEM}) with (\ref{hQck/}) represents a combination of the energy and momentum balances of the matter fluid. Combining (\ref{BEM}) and (\ref{hQck/}) provides us with
\begin{equation}
\dot{\delta}_{M}^{c} + \frac{8\pi G}{3} \sigma \delta_{M}^{c}  + K\hat{\Theta}^{c} = 0\
\ \quad \Rightarrow\quad \hat{\Theta}^{c} = - \frac{1}{K}\left[\dot{\delta}_{M}^{c} + \frac{8\pi G}{3} \sigma \delta_{M}^{c}\right]\ ,
\label{ddc}
\end{equation}
where
\begin{equation}
K \equiv 1 + \frac{\sigma H}{3\rho_{M}}\left(1 - \frac{4\pi G}{3}\frac{\sigma}{H} - \frac{\sigma H}{3\rho_{M}}\,\frac{k^{2}}{a^{2}H^{2}}\right)
\ .
\label{K}
\end{equation}
In the next step we
differentiate (\ref{ddc}), eliminate $\dot{\hat{\Theta}}\,^{c}$ through the perturbed Raychaudhuri equation and  $\hat{\Theta}^{c}$ through (\ref{ddc}) and change to the scale factor $a$ as independent variable.
The result is the following closed second-order equation for $\delta_{M}^{c}(a)$:
\begin{equation}
\delta_M^{c\prime\prime} + g(a)\delta_M^{c\prime} + f(a)\delta_M^{c} = 0
\ ,
\label{fineq}
\end{equation}
where the prime means a derivative with respect to $a$.
The coefficients $g(a)$ and $f(a)$ are explicitly known and depend only on the background dynamics. They are given by
\begin{equation}
g(a) = \frac{1}{a}\left[\frac{3}{2} + 3B - \frac{L}{K} + \frac{1}{3}A\frac{k^{2}}{H^{2}a^{2}}\right]
\
\label{g1}
\end{equation}
and
\begin{equation}
f(a) = -\frac{1}{a^{2}}\left[\left(\frac{3}{2} - \frac{3}{2}B\right)K
+ B\left(\frac{L}{K} - 2 - \frac{1}{2}B - \frac{1}{3}A\frac{k^{2}}{a^{2}H^{2}}\right)\right]
\ ,
\label{f1}
\end{equation}
with
\begin{equation}
A \equiv \frac{\sigma H}{\rho_{M}} = \frac{1 - \Omega_{M0}}{\ \Omega_{M0}a^{-3/2}}
\ ,
\label{sH/r}
\end{equation}
\begin{equation}
B \equiv \frac{8\pi G}{3}\frac{\sigma}{H} = \frac{1 - \Omega_{M0}}{1 - \Omega_{M0} + \Omega_{M0}a^{-3/2}}
\ ,
\label{s/H}
\end{equation}
\begin{equation}
K = 1 + \frac{1}{3}A - \frac{1}{6}AB - \frac{1}{9}A^{2}\,\frac{k^{2}}{a^{2}H^{2}}
\ ,
\label{K1}
\end{equation}
and
\begin{equation}
L = \frac{1}{2}B + \frac{1}{4}B^{2}A + \frac{1}{9}A^{2}\left(3B-4\right)\frac{k^{2}}{a^{2}H^{2}}
\ .
\label{L1}
\end{equation}
Equation (\ref{fineq}) for $\delta_{M}^{c}$ encodes the entire perturbation dynamics of the model, it is the central relation of the paper.
The perturbation $\delta_{X}^{c} \equiv \frac{\hat{\rho}_{X}^{c}}{\rho_{X}}$ of the dark energy can be obtained from $\hat{\rho}_{X}^{c} = \frac{\sigma}{3}\hat{\Theta}^{c}$ with $\hat{\Theta}^{c}$ from
(\ref{ddc}). It is determined by $\delta_{M}^{c}$ and its first derivative:
\begin{equation}
\delta_{X}^{c} = - \frac{1}{3K}\left[a\delta_{M}^{c\prime}
+ B\delta_{M}^{c}\right]\ .
\label{deltax}
\end{equation}
The coefficients $A$ and $B$ do not depend on the wavenumber $k$, they are of the order of one around the present time with $a \approx 1$. The quantity $K$, however, is scale-dependent. On scales well inside the
horizon, equivalent to  $\frac{k^{2}}{a^{2}} \gg H^{2}$, one has $|K|\gg 1$. This suggests that, on scales which are relevant for structure formation, the fluctuations $\delta_{X}^{c}$ of the dark energy should be very small compared with the matter fluctuations $\delta_{M}^{c}$. The quantitative analysis of section
\ref{The power spectrum} below will confirm this behavior.
It is expedient to note that Eq.~(\ref{fineq}) is of the same structure as the corresponding perturbation equation for a unified viscous dark sector fluid \cite{VDF}. However, the coefficients $g(a)$ and $f(a)$ are different. In particular, the wave-number dependence is more complicated for the present model.

\subsection{Non-adiabatic perturbations}

To clarify the non-adiabatic character of the model it is useful to consider the gauge-invariant perturbations of the dark pressure, $\hat{p}^{c}_{X} \equiv \hat{p}_{X} + \dot{p}_{X} v$ and of the dark energy, $\hat{\rho}^{c}_{X} \equiv \hat{\rho}_{X} + \dot{\rho}_{X} v$.
Let us assume that in the corresponding rest frame
$\hat{p}^{c}_{X} = c_{s}^{2}\hat{\rho}^{c}_{X}$.
Then, $\hat{p}_{X}$ may generally be written as
\begin{equation}
\hat{p}_{X} = c_{s}^{2}\hat{\rho}_{X} + \left(c_{s}^{2} -
\frac{\dot{p}_{X}}{\dot{\rho}_{X}}\right)\,v\,\dot{\rho}_{X}\ .\label{hpx}
\end{equation}
In the general case, the non-adiabatic part of the perturbations of the $X$ component is given by
\begin{equation}
\hat{p}_{X} - \frac{\dot{p}_{X}}{\dot{\rho}_{X}}\hat{\rho}_{X}
= \left(c_{s}^{2} -
\frac{\dot{p}_{X}}{\dot{\rho}_{X}}\right)\,\hat{\rho}^{c}_{X}\ .\label{nadx}
\end{equation}
For the present model $c_{s}^{2} = \frac{\dot{p}_{X}}{\dot{\rho}_{X}} = -1$ is valid, i.e., the $X$ component by itself is adiabatic.
For  the total non-adiabatic perturbations we have
\begin{equation}
\hat{p}- \frac{\dot{p}}{\dot{\rho}}\hat{\rho} = \hat{p}_{X} -\frac{\dot{p}_{X}}{\dot{\rho}}\,\hat{\rho}
= \hat{p}_{X} + \frac{\dot{\rho}_{X}}{\dot{\rho}}\left(\hat{\rho}_{M} + \hat{\rho}_{X}\right)\ .\label{nadtot}
\end{equation}
Introducing here (\ref{hpx}) and rearranging we obtain
\begin{equation}
\hat{p}_{nad} \equiv\hat{p}- \frac{\dot{p}}{\dot{\rho}}\hat{\rho} = \frac{\dot{\rho}_{X}\dot{\rho}_{M}}{\dot{\rho}}
\left(\frac{\hat{\rho}_{M}}{\dot{\rho}_{M}} - \frac{\hat{\rho}_{X}}{\dot{\rho}_{X}}\right) +
\left(c_{s}^{2} + 1\right)\hat{\rho}_{X}^{c}\ .\label{nadtot1}
\end{equation}
Although the $X$ component has no intrinsic non-adiabaticity for
the present case, $c_{s}^{2} = -1$ and
$\hat{p}_{X} = - \hat{\rho}_{X}$, the two-component system as a whole is non-adiabatic.
As it is obvious from (\ref{nadtot1}), non-adiabatic first-order perturbations can  be characterized by the quantity
\begin{equation}
\frac{\hat{\rho}_{M}}{\dot{\rho}_{M}} - \frac{\hat{\rho}_{X}}{\dot{\rho}_{X}} \neq 0
.\label{nad}
\end{equation}
Since
\begin{equation}
\frac{\hat{\rho}_{M}^{c}}{\dot{\rho}_{M}} = \frac{\delta_{M}^{c}}{-3H + 4\pi G\sigma}\  \quad \mathrm{and }\quad
 \frac{\hat{\rho}_{X}^{c}}{\dot{\rho}_{X}} = - \frac{\hat{\Theta}^{c}}{12 \pi G \rho_{M}}
\ ,\label{frc}
\end{equation}
where we have used (\ref{drmq}) and (\ref{drxq}), respectively, it follows with (\ref{ddc}) that
\begin{equation} \frac{\hat{\rho}_{M}^{c}}{\dot{\rho}_{M}} - \frac{\hat{\rho}_{X}^{c}}{\dot{\rho}_{X}} =
\frac{\delta_{M}^{c}}{-3H\left[1 - \frac{4\pi G\sigma}{3H}\right]}
- \frac{2}{3}\frac{\rho_{0}}{3H_{0}^{2}\rho_{M}K}\left[\delta_{M}^{c\prime}aH + H_{0}\left(1-\Omega_{M0}\right)\delta_{M}^{c}\right]
.\label{nadc}
\end{equation}
Notice that
\begin{equation}
\frac{\hat{\rho}_{M}}{\dot{\rho}_{M}} - \frac{\hat{\rho}_{X}}{\dot{\rho}_{X}} = \frac{\hat{\rho}_{M}^{c}}{\dot{\rho}_{M}} - \frac{\hat{\rho}_{X}^{c}}{\dot{\rho}_{X}}
, \label{inv}
\end{equation}
i.e., the combination  (\ref{nad}) is gauge-invariant, although the single terms by themselves are not.
Equation (\ref{nadc}) implies that the non-adiabatic contribution is completely determined by $\delta_{M}^{c}$ and its first derivative. Once $\delta_{M}^{c}(a)$ is known, the expression (\ref{nadc}) is determined as well. The existence of non-adiabatic perturbations is characteristic for interacting two-component systems. In our case, the non-interacting limit corresponds to $\sigma =0\ \Rightarrow\ \rho_{X} = 0\ \Rightarrow\ \Omega_{M0} = 1$, equivalent to the one-component Einstein - de Sitter universe which, of course, has a purely adiabatic perturbation dynamics.

\subsection{Almost adiabatic initial conditions}
\label{adiabatic}

For $a \ll 1$, the non-adiabaticity (\ref{nadc}) can be calculated explicitly. This is relevant for specifying
the initial conditions. It will turn out that, for the present model, almost adiabatic initial conditions are appropriate.
At high redshifts, i.e., for $a\ll 1$, Eq. (\ref{fineq}) reduces to
\begin{equation}\label{early}
\delta^{c\prime\prime}_{M} + \frac{3}{2a} \,\delta^{c\prime}_{M} - \frac{3}{2a^2}\,\delta^{c}_M =0 \  \qquad \qquad  (a \ll 1)\ ,
\end{equation}
with solutions
\begin{equation}
\delta_{M}^{c} = c_1 a + c_2 a^{-3/2} \ \quad \Rightarrow\quad \delta^{c\prime}_{M} = c_1 - \frac{3}{2}c_2 a^{-5/2} \qquad \qquad  (a \ll 1)\ .
\label{asympsol}
\end{equation}
The asymptotic values for $H$, $\rho_{M}$ and $K$ for $a \ll 1$ are
\begin{equation}
H \approx H_{0}\Omega_{M0}a^{-3/2}\ ,\quad \rho_{M} \approx \rho_{0}\Omega_{M0}^{2}a^{-3}\ ,\quad K\approx1\  \qquad \qquad  (a \ll 1)\ .
\label{HrK}
\end{equation}
Applying (\ref{asympsol}) and (\ref{HrK}) in (\ref{nadc}) we find, for $a \ll 1$,
\begin{eqnarray}
\frac{\hat{\rho}_{M}^{c}}{\dot{\rho}_{M}} - \frac{\hat{\rho}_{X}^{c}}{\dot{\rho}_{X}} &\approx&
- \frac{a^{3/2}}{3H_{0}\Omega_{M0}}\left[c_1 a + c_2 a^{-3/2}\right] \nonumber\\
&&- \frac{2}{9}\frac{a^{3}}{H_{0}^{2}\Omega_{M0}^{2}}\left[\left(c_1 - \frac{3}{2}c_2 a^{-5/2}\right)a^{-1/2}H_{0}\Omega_{M0}
+ H_{0}\left(1 - \Omega_{M0}\right)\left(c_1 a + c_2 a^{-3/2}\right)\right]
\ .\nonumber\\\label{nad<}
\end{eqnarray}
The $a$-independent $c_{2}$-terms cancel each other exactly. The leading term for the growing mode is
\begin{equation}
 \frac{\hat{\rho}_{M}^{c}}{\dot{\rho}_{M}} - \frac{\hat{\rho}_{X}^{c}}{\dot{\rho}_{X}} \approx
 - \frac{5}{9}\frac{c_{1}}{H_{0}\Omega_{M0}}\,a^{5/2}
 \qquad \qquad  (a \ll 1)\ .
\label{nad0}
\end{equation}
It is interesting to note, that both terms on the left-hand side contribute with comparable amounts, although
$\hat{\rho}_{M}^{c} \gg \hat{\rho}_{X}^{c}$.
The non-adiabatic pressure perturbation (\ref{nadtot1}) associated with the growing mode becomes
\begin{equation}
\hat{p}_{nad}  \approx \frac{5}{6}c_{1}\frac{\rho_{0}\Omega_{M0}\left(1 - \Omega_{M0}\right)}{a^{1/2}}\qquad \qquad  (a \ll 1)\ ,\label{nadtot1a<<}
\end{equation}
the corresponding fractional quantity is
\begin{equation}
\frac{\hat{p}_{nad}}{\rho + p}  \approx \frac{5}{6} \frac{1 - \Omega_{M0}}{\Omega_{M0}}c_{1}a^{5/2}\qquad \qquad  (a \ll 1)\ .\label{nadfrac}
\end{equation}
It is only for $\Omega_{M0}=1$, corresponding to the one-component Einstein-de Sitter model, that
the non-adiabatic contribution exactly vanishes. Although the perturbation $\delta_{X}^{c}$ in (\ref{deltax}) is different from zero for $\Omega_{M0} = 1$, it does not contribute to the total energy density perturbation
since the background density of the $X$ component vanishes (see relation (\ref{tot}) below). The ratio
\begin{equation}
\frac{|\hat{\rho}_{X}|}{|\hat{\rho}_{M}|}  \approx \frac{1}{3} \frac{1 - \Omega_{M0}}{\Omega_{M0}}a^{3/2}\qquad \qquad  (a \ll 1)\ \label{ratiopert}
\end{equation}
of the perturbations scales as the background ratio for these components (cf. Eqs.~(\ref{rmom}) and (\ref{rxom})).
As expected, the non-adiabatic pressure perturbation is of the order of the dark-energy perturbation:
\begin{equation}
\frac{|\hat{p}_{nad}|}{|\hat{\rho}_{X}|}  \approx \frac{5}{2} \qquad \qquad  (a \ll 1)\ .\label{rationad}
\end{equation}
It follows that at high redshifts it is not only the background contribution of the dark energy density which is small compared with the total background energy density, the dark-energy perturbations, together with the non-adiabatic pressure perturbations, are much smaller than the total energy-perturbations as well.

On the other hand,  Eq.~(\ref{early}) is also the correct one-component limit of Eq.~(\ref{fineq}) for $\Omega_{M0} = 1$, i.e., the Einstein-de Sitter model  with a purely adiabatic perturbation dynamics. Consequently, the smallness of the non-adiabatic contribution at high redshift allows us to use approximately adiabatic initial conditions for $k>0$.
However, the mere two-component nature of our model requires that there is always a small, non-vanishing admixture of non-adiabaticity. This is consistent with the result in \cite{Wands}, that purely adiabatic large-scale perturbations can never give rise to entropy perturbations.

\section{Dark energy perturbations and power spectrum}
\label{The power spectrum}

\begin{center}
\begin{figure}[t!]
\begin{minipage}[t]{0.3\linewidth}
\includegraphics[width=\linewidth]{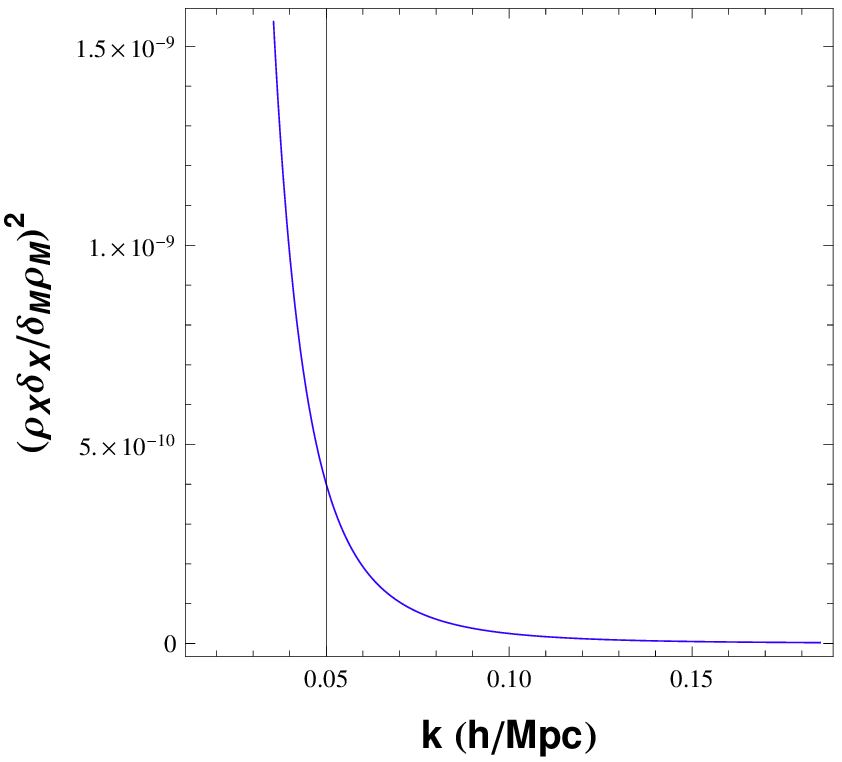}
\end{minipage}
\begin{minipage}[t]{0.3\linewidth}
\includegraphics[width=\linewidth]{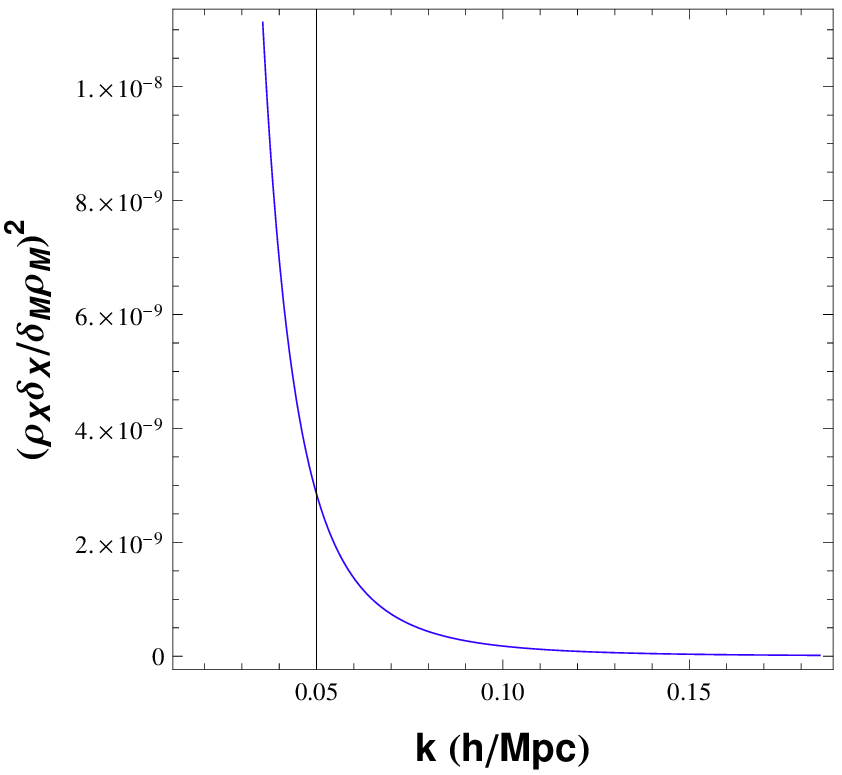}
\end{minipage} \hfill
\caption{{\protect\footnotesize Relative power spectrum (\ref{relX}) as a function of $k$ for $\Omega_{M0} = 0.3$ (left) and $\Omega_{M0} = 0.8$ (right).}}
\label{dXdM}
\end{figure}
\end{center}

\begin{center}
\begin{figure}[t!]
\begin{minipage}[t]{0.3\linewidth}
\includegraphics[width=\linewidth]{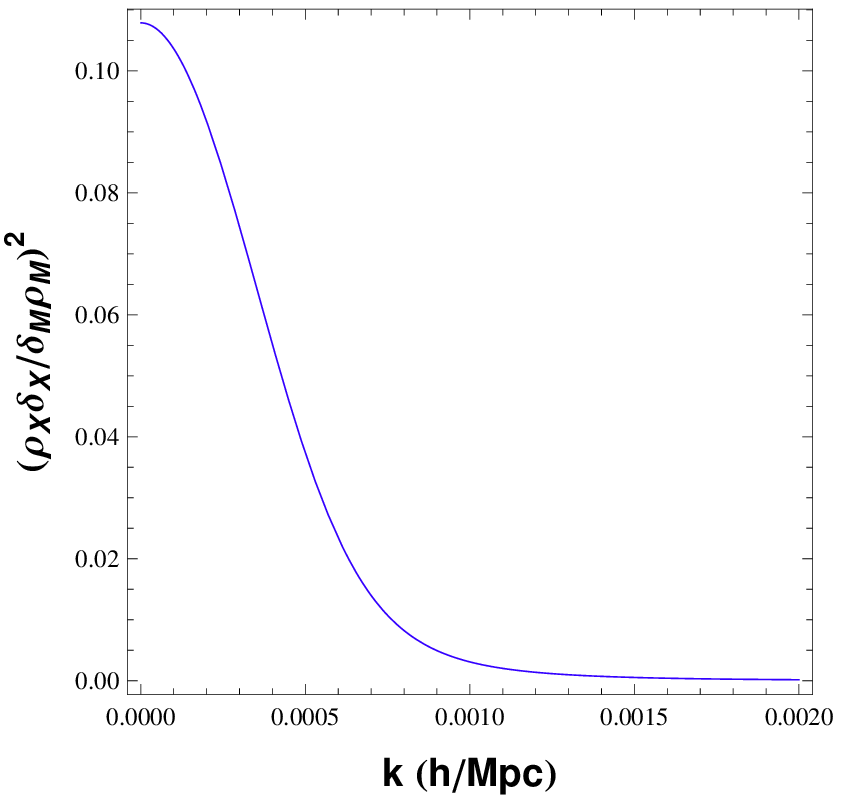}
\end{minipage}
\begin{minipage}[t]{0.3\linewidth}
\includegraphics[width=\linewidth]{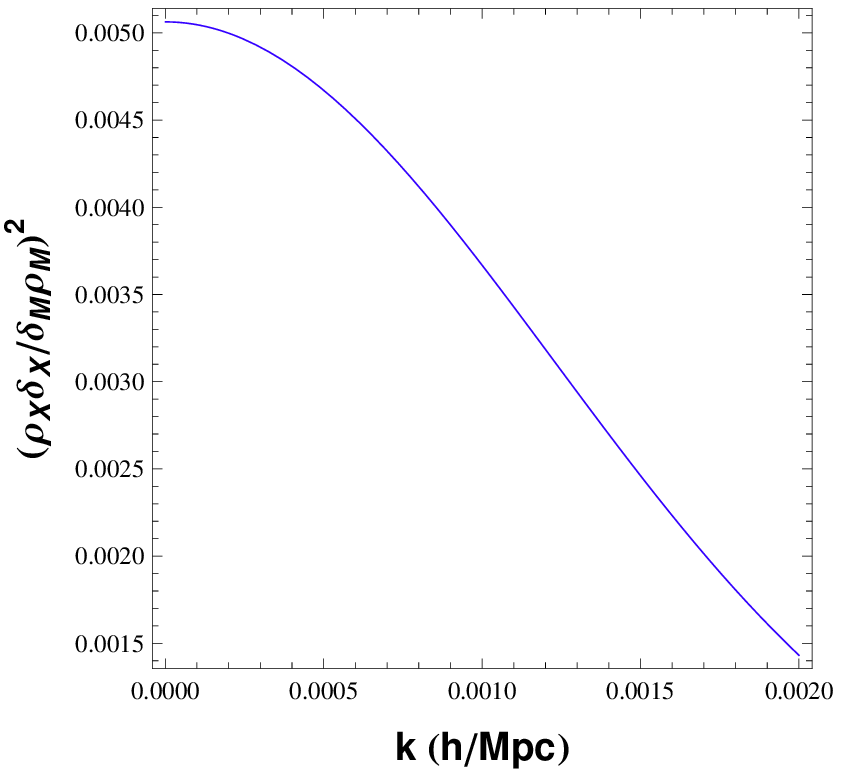}
\end{minipage} \hfill
\caption{{\protect\footnotesize Relative power spectrum (\ref{relX}) as a function of $k$ on large scales for $\Omega_{M0} = 0.3$ (left) and $\Omega_{M0} = 0.8$ (right).}}
\label{LS}
\end{figure}
\end{center}

\begin{center}
\begin{figure}[t!]
\begin{minipage}[t]{0.3\linewidth}
\includegraphics[width=\linewidth]{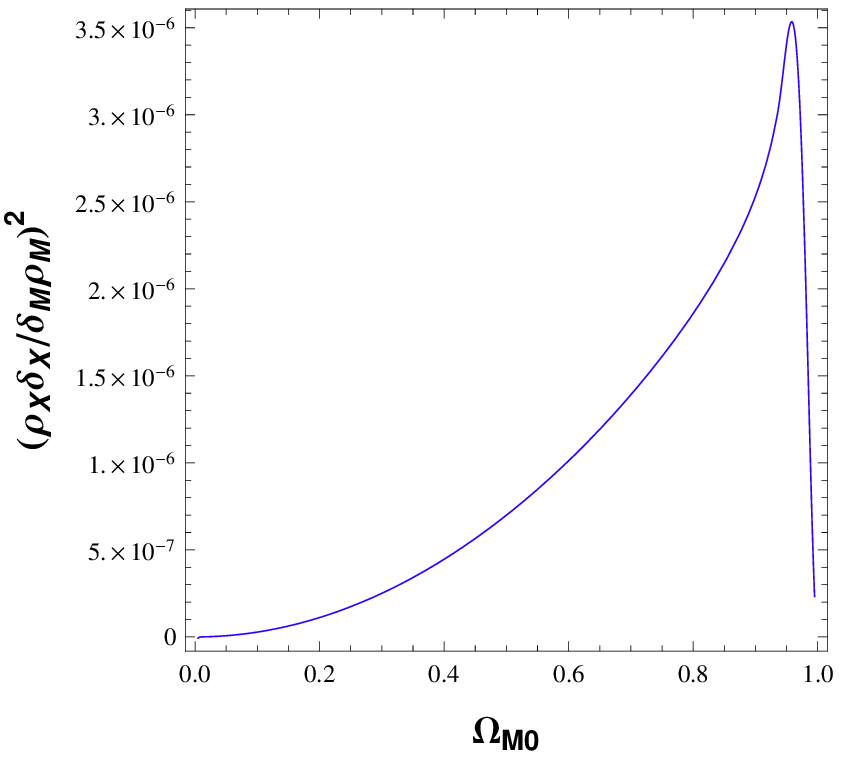}
\end{minipage}
\begin{minipage}[t]{0.3\linewidth}
\includegraphics[width=\linewidth]{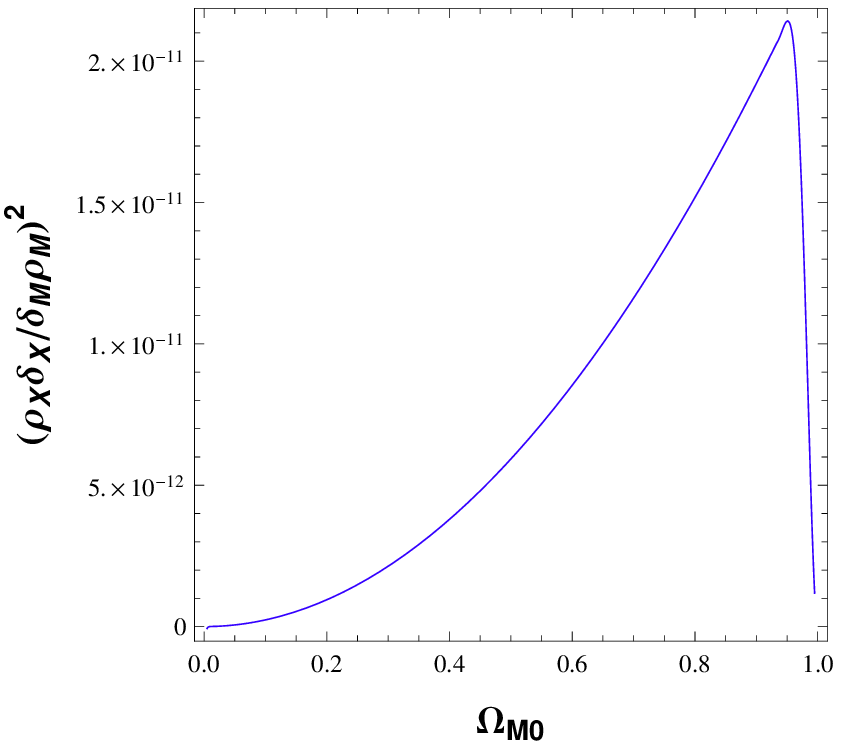}
\end{minipage} \hfill
\caption{{\protect\footnotesize Relative power spectrum (\ref{relX}) as a function of $\Omega_{M0}$ for $k = 0.01$ (left) and $k = 0.185$ (right).}}
\label{dXdM2}
\end{figure}
\end{center}

\begin{center}
\begin{figure}[t!]
\begin{minipage}[t]{0.3\linewidth}
\includegraphics[width=\linewidth]{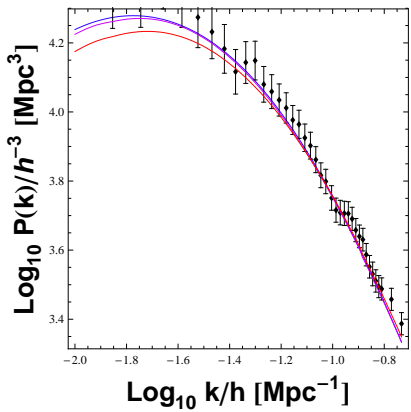}
\end{minipage} \hfill
\caption{{\protect\footnotesize The best fit of the observed 2dFGRS power spectrum. The lines correspond to the BBKS transfer function (blue), the approximate $\Lambda$CDM best fit using the numerical method of \cite{Julio} (red) and the corresponding fit of the interacting model (violet).}}
\label{spectrum}
\end{figure}
\end{center}

In the present study we have neglected baryons. This can be considered a reasonable approximation since baryons represent only $5\%$ of the total amount of energy.
Nevertheless one should keep in mind that the observed power spectrum reflects the distribution of luminous, baryonic matter. For the $\Lambda$CDM model with a ``true" cosmological constant, the distribution of baryons can approximately be determined by the dark matter distribution, i.e., the observed power spectrum is proportional to $\delta_M^2$. In dynamic dark-energy models as in the present case, however, the gravitational potential - which determines the baryonic distribution - is not only due to dark matter perturbations, but also due to the perturbations in the cosmological term.
The total energy density perturbation $\delta^{c}$ is given by
\begin{equation}
\delta^{c} \equiv \frac{\hat{\rho}_{M}^{c} + \hat{\rho}_{X}^{c}}{\rho} = \frac{\rho_{M}}{\rho}\delta^{c}_{M} + \frac{\rho_{X}}{\rho}\delta^{c}_{X}
\ .
\label{tot}
\end{equation}
Only if the $\delta_X^{c}$ term is negligible, the observed spectrum can be related with $\delta_M^{2}$.
The relative importance of the dark-energy fluctuations can be quantified by the expression
\begin{equation}
\frac{\rho_{X}}{\rho_{M}}\frac{\delta^{c}_{X}}{\delta^{c}_{M}} = - \frac{1}{3K}\frac{1-\Omega_{M0}}{\Omega_{M0}}a^{3/2}\left[\frac{d \ln \delta^{c}_{M}}{d \ln a} + B\right]
\ ,
\label{relX}
\end{equation}
where we have applied equations (\ref{ratioom}) and (\ref{deltax}).
In the integration of equation (\ref{fineq}) we have used scale-invariant, adiabatic initial conditions for $k>0$ at $z = 1000$, which is a very good approximation as discussed in subsection \ref{adiabatic} above. In Figure \ref{dXdM} we show the ratio (\ref{relX}) at present as a function of $k$ for the range $0.01\,Mpc^{-1}h < k < 0.2\,Mpc^{-1}h$ for two different values of $\Omega_{M0}$.
This ratio is much smaller than $1$ in the entire region.
Hence, we can see that perturbations in the vacuum term are actually negligible. The fact that (\ref{relX}) decreases with $k$ demonstrates that the dark energy component is perturbed significantly at the most at very large scales $k \sim 0$.
But as Figure \ref{LS} reveals, the dark-energy perturbations remain smaller than the matter perturbations even on the largest scales. In Figure \ref{dXdM2} we also show the ratio  (\ref{relX}) as a function of $\Omega_{M0}$, for two fixed scales. As to be expected, vacuum perturbations go to zero in the limits $\Omega_{M0} = 0$ (de Sitter solution) and $\Omega_{M0} = 1$ (Einstein-de Sitter case).
The result for larger $k$ can be seen as a justification of the analysis
performed in \cite{Julio} with $\hat{\rho}_{X} = 0$. As already mentioned, a
good fit of the observed spectrum was found in \cite{Julio} for a present
matter density parameter around $0.48$. The corresponding best fit is
shown in Figure \ref{spectrum}, where the spectrum was normalized with the
BBKS transfer function \cite{BBKS} for large $k$.

It should be mentioned that the results of the present analysis are not in accordance with those presented by some of us in \cite{[9]}, where the vacuum term was perturbed and a fractional matter density near $1$ was obtained. Some of the reasons for this discrepancy are the following. i) As initial condition, the $\Lambda$CDM-model based BBKS power spectrum at the redshift of last scattering has been used in \cite{[9]}, something which is not completely justifiable since the present model is different from the $\Lambda$CDM model. ii) Only the conservation equations for the total fluid were considered in \cite{[9]}, but not the separate balance equations for the individual components. In this way, the non-adiabaticity of the system was not explicitly explored. iii) A separately conserved baryon component with a fixed relative density was included in \cite{[9]}. For this reason, the dark-matter density parameter could not vary in the entire interval [0,1], which led to some numerical difficulties. In the present gauge invariant treatment on the other hand, the non-adiabatic perturbations appear explicitly and scale-invariant initial conditions are implemented at very high redshifts. Our result corroborates the prevailing belief that in a fluid with equation-of-state parameter $\omega = -1$ no significant perturbations should be expected on scales inside the horizon.

As a final point we mention that for negligible dark energy perturbations $\hat{\rho}_X \approx 0$ the relative entropy perturbation $\hat{p}_{nad}$ in (\ref{nadtot1}) with $c^2=-1$ reduces to
\begin{equation}
\frac{\hat{p}_{nad}}{\hat{\rho}_M} \approx \frac{\dot{\rho}_X}{\dot{\rho}}  = \frac{1}{2}\frac{1-\Omega_{M0}}{1-\Omega_{M0} + \Omega_{M0}a^{-3/2}}\ .
\end{equation}
Obviously, the entropic perturbations increase with the scale factor from a very small value for $a\ll 1$ to $1/2$ in the long-time limit $a\gg 1$.
The present value is $(1-\Omega_{M0})/2$, leading to approximately $25\%$ of non-adiabaticity.

\section{Conclusions}
\label{Conclusions}

We have investigated an interacting dark-energy model in which a vacuum term decays into dark matter linearly
with the Hubble rate. The homogeneous and isotropic background evolution of this model is given by simple, explicitly known functions of the scale factor. The two-component linear scalar perturbation dynamics was reduced to a single, second-order equation for the gauge-invariantly defined, comoving fractional density contrast of the dark matter component. Thanks to the specific coupling between the components, the dark-energy perturbation is  determined by a combination of the dark-matter perturbation and its first time derivative. The  presence of relative entropy perturbations makes the entire dynamics intrinsically non-adiabatic. However, as we demonstrated analytically, this non-adiabaticity is very small at early times. Consequently, for our numerical analysis adiabatic initial conditions could be used as a good approximation for $k>0$ at high redshifts. As a main result of this analysis we obtained that dark-energy fluctuations are negligible on scales that are relevant for cosmic structure formation. In other words, for our model dark energy does not cluster on small scales. This result provides an \textit{a posteriori} justification of the analysis in \cite{Julio} where dark-energy perturbations were neglected by assumption. This seems also to imply the conclusion obtained in \cite{Julio}, that the
matter-power spectrum is only consistent with a present matter-density
parameter $\Omega_{M0} \approx 0.48$. This value is compatible (within
$2\sigma$ confidence level) with an updated joint background analysis of the
$\Lambda\propto H$ model in \cite{JailsonIII}. On scales larger than the
horizon the contribution of dark-energy fluctuation increases but remains
smaller than the dark-matter contribution even in the limit $k\rightarrow
0$.

\section*{Acknowledgements}

This work was partially supported by CNPq, CAPES, FAPES and FAPESB.


{}

\end{document}